\documentclass[12pt]{article}
\usepackage{amsfonts,epsfig,latexsym,amscd}

\newcommand{\CP}{{\mathbb{C}}{{P}}}

\newcommand{\bea}{\begin{eqnarray}}
\newcommand{\eea}{\end{eqnarray}}
\newcommand{\pr}{\partial}

\newcommand{\Tr}{{\rm Tr}}

\newcommand{\be}{\begin{equation}}
\newcommand{\ee}{\end{equation}}

\def\E {{\cal E}}

\def\pr {\partial}
\def\half {\frac{1}{2}}

\begin{document}
\begin{titlepage}
\title{\vskip -70pt
\begin{flushright}
{\normalsize DAMTP-2004-91} \\
\end{flushright}\vskip 50pt
{\bf The Energy of Scattering Solitons in the Ward Model}
}
\vspace{1cm}
\author{{T. Ioannidou}\thanks{e-mail address: ti3@auth.gr}\\
{\sl Mathematics Division, School of Technology}\\
{\sl Aristotle University of Thessaloniki} \\
{\sl Thessaloniki 54124, Greece} \\  
and\\
{N. S. Manton}\thanks{e-mail address: N.S.Manton@damtp.cam.ac.uk}\\
{\sl Department of Applied Mathematics and Theoretical Physics}\\
{\sl University of Cambridge} \\
{\sl Wilberforce Road, Cambridge CB3 0WA, England}\\ \\
}
\date{September, 2004}
\maketitle
\thispagestyle{empty}
\begin{abstract}
\noindent
The energy density of a scattering soliton solution in Ward's integrable chiral
model is shown to be instantaneously the same as the energy density of
a static multi-lump solution of the $\CP^3$ sigma model. This explains
the quantization of the total energy in the Ward model. 
\end{abstract}
\end{titlepage}

\section{Introduction}
\label{sec:intro}

The Ward model (Ward 1988) is one of the most interesting models of soliton
dynamics. It is defined in two space dimensions and is almost fully
Lorentz invariant. The field equation can be regarded as the
(2+1)-dimensional Bogomolny equation for Yang-Mills-Higgs fields, with
gauge group SU(2). This formulation is first order in time derivatives
and is fully Lorentz invariant, but there is no natural Lagrangian and 
hence Noether's theorem is not available to construct conservation laws.

An alternative formulation, due to Ward, involves fixing the gauge,
and this requires a choice of spatial direction. It leads to a field
equation for an SU(2)-valued ``chiral'' field $J$ which is second  
order in time derivatives,
\be
\pr_t(J^{-1}\pr_t J) - \pr_x(J^{-1}\pr_x J) - \pr_y(J^{-1}\pr_y J) +
   [J^{-1}\pr_y J, J^{-1}\pr_t J] = 0 \,.
\ee
This equation is a variant of that of the standard chiral sigma model 
in 2+1 dimensions for $J$, where the final commutator term is absent. In this
formulation, Lorentz invariance is broken to (1+1)-dimensional Lorentz 
invariance. However, there is a Lagrangian, and hence a conserved
energy and a conserved momentum in the $y$-direction. One might imagine
that by unfixing the gauge one could recover a complete conserved
energy and momentum vector, but it is not clear how this should be done.

A remarkable feature of the Ward model is that it is integrable, both
from the point of view of the (2+1)-dimensional Bogomolny equation and 
from the point of view of the equation for the chiral field $J$. The
integrability ultimately arises from the fact that the Bogomolny
equation is a reduction of the (2+2)-dimensional self-dual Yang-Mills
equation, which can be treated by twistor methods. Because of the
integrability there is a holomorphic structure, an infinite set of
conservation laws (Ioannidou \& Ward 1995), including the energy and momentum
mentioned above, 
and a procedure for obtaining general solutions in terms of initial
data using spectral methods and via the solution of a Riemann-Hilbert
problem (Fokas \& Ioannidou 2001). For us, the most useful consequence 
of integrability is that
there are families of algebraically explicit solutions representing
multi-soliton scattering processes.

These soliton solutions are constructed from auxiliary functions of a
spectral parameter $\lambda$, with poles. If there are a finite number
of simple poles then the solution describes a finite number of solitons
moving along straight lines in the plane at constant speed, and they
do not scatter.

Ward (1995) noticed that if there is a double pole in $\lambda$
then the solution represents scattering solitons. The double pole can
be obtained (with care) in the limit that simple poles coalesce. The
construction is technically a bit more difficult than the simple pole
case. It is related to the construction of static solutions of
the planar chiral sigma model, in terms of so-called unitons
(Uhlenbeck 1989). Whereas
the simple pole solutions correspond to single unitons, the double
pole solutions correspond to 2-unitons. The original example of a
double pole solution was generalised by Ioannidou (1996), and a number of
interesting multi-soliton scattering solutions were found.
The general solution of this type depends on an arbitrarily
chosen pair of rational functions $f(z)$ and $h(z)$, where $z=x+iy$ is the
complex coordinate in the spatial plane. 
Recently, a very general analysis of multi-pole,
multi-soliton solutions has been given (Dai \& Terng 2004).

In this paper we shall make some observations concerning the
energy density and total energy of the general double pole solution
discussed by Ward (1995) and Ioannidou (1996), which has no net momentum. 
Our proofs of these observations are not deep, and we expect
that a better understanding is possible.

Our first observation is that the total energy of the double pole
solutions is quantized in units of $8\pi$. This was actually noted
earlier by Ioannidou \& Zakrzewski (1998) in a few examples, but no proof
was given. It is physically quite a surprising result, because the
solutions represent solitons in motion and part of the energy is
kinetic. We are familiar with quantized energy for multi-solitons
at rest, if they satisfy a Bogomolny equation, but usually any kinetic 
energy raises the total energy above the quantized value 
(see e.g. Manton \& Sutcliffe 2004). We shall clarify below the number 
of energy units that a particular solution has  -- the relation to the 
number of solitons is not particularly simple.

Our second observation is that the energy density of the solutions
(including both the gradient and kinetic energies of the field),
calculated at any instant of time $t$, is the same as the energy
density of a static $\CP^3$ sigma model solution (also known as a
$\CP^3$ multi-lump). The $\CP^3$ multi-lump has $t$ as a
parameter, but is not regarded as dynamical and has no kinetic
energy. We have no deep understanding of this relationship between the
SU(2) Ward model solutions and the $\CP^3$ multi-lumps; we were led to
it through the algebraic form of the energy density in a number of
examples. Our proof that the energy density is the same is based on
direct calculation.

The first observation follows from the second. It is well known that
the total energy of a static $\CP^3$ multi-lump solution is quantized
(Zakrzewski 1989),
and we can determine how many units of energy there are in terms of
the degrees of $h$, $f$, and the $z$-derivative of $f$.

\section{Double Pole Solutions of the Ward Model}
\setcounter{equation}{0}
\label{sec:double}

The simplest solutions of the Ward model arise from a function of
$\lambda$ with a single pole $\frac{R(z,\bar{z})}{\lambda -\mu}$, where $\mu$
is a constant, and $R(z,\bar{z})$ is a $2 \times 2$ matrix depending on a
rational function $f(z)$. For $\mu=i$ this solution is static and is equivalent
to the multi-lump solution of the $\CP^1$ sigma model given by the
rational map $f(z)$. The topological charge $N$ of the lump is the
degree of $f$, and the energy in the Ward model is the same as in the
sigma model, namely $8\pi N$.

The next simplest solutions in the Ward model are those based on a
superposition of simple poles in the $\lambda$-plane. Let us denote
the pole locations by $\mu_1, \mu_2, \dots, \mu_n$, which are
necessarily distinct. Associated with
these are rational functions $f_1(z), f_2(z), \dots, f_n(z)$. The
function $f_r(z)$ describes a multi-lump which moves at a constant
velocity determined by $\mu_r$. Since the pole locations are distinct,
so are the velocities, so at most one of the multi-lumps is at
rest. Although the multi-lumps are in motion, they do not scatter off
each other in this type of solution. There is a simple formula for the total
energy, which varies continuously as the parameters $\mu_r$ vary, so
the energy is not quantized.

Scattering soliton solutions are obtained by considering a limit of
the solutions above, in which two or more of the pole locations
$\mu_r$ are brought into coincidence. The simplest case is where $n=2$
and the parameters $\mu_1$ and $\mu_2$ are brought into coincidence at
$i$. If the limit is taken appropriately, a double pole at $\lambda =
i$ results, and it is in this case that the momentum in the $y$-direction 
vanishes and the energy is quantized as a
multiple of $8\pi$. Solutions with a double pole at a different
location are Lorentz boosted and have a different energy.

It was shown by Ward (1995) that for these double pole solutions the
SU(2) matrix $J$ has a factorized 2-uniton form. We recall now the
expression for $J$. It depends on two rational functions $f(z)$ and
$h(z)$ (which arise from the functions $f_1(z)$ and $f_2(z)$ as the limit
of coalescing poles is taken). Explicitly
\be
\label{2Unit}
J = \left( I - 2\,\frac{q_1^\dag \otimes q_1}{|q_1|^2} \right)
\left( I - 2\,\frac{q_2^\dag \otimes q_2}{|q_2|^2} \right),
\ee
where $I$ is the $2\times 2$ unit matrix, and $q_1$ and $q_2$ are the
2-component row vectors
\begin{eqnarray}
q_1 &=& (1+|f|^2)(1,f) - 2i(tf'+h)(\bar{f},-1), \label{q1}\\
q_2 &=& (1,f)  \label{q2} \,.
\end{eqnarray}
$q_2$ is just a function of $z$, but $q_1$ and hence $J$ are functions
of $z$, $\bar{z}$ and $t$. $f'$ denotes $\frac{df}{dz}$. Because of the
time dependence of $q_1$, the solution describes the scattering of solitons.

Ioannidou (1996) explored a number of examples of the 2-uniton solution 
(\ref{2Unit}). The simplest exhibiting soliton scattering is the 
solution with $f(z)=z$ and $h(z)=z^2$ (Ward 1995). The scattering can be
deduced from the zeros of the expression $tf'+h=t+z^2$, noting that these
are located at $z=\pm\sqrt{-t}$. The zeros are on the $x$-axis for $t<0$, pass
through the origin at $t=0$, and are on the $y$-axis for $t>0$.
A slightly more complicated example is with  $f(z)=z$ and
$h(z)=z^3$. Here three solitons scatter with equilateral triangular
symmetry. Solutions with $h=0$ are also possible. If $f(z)=z$ and
$h=0$, for example, then there is circular symmetry for all $t$, with
a ring-like soliton contracting and then expanding.

It is rather curious that these solutions (and also their energy
density, which we investigate in the next section) appear to respect the
rotational symmetry of the plane, despite our earlier remarks about
how Lorentz invariance is partially broken. It is possible that these
solutions have zero momentum and the same energy no matter which
direction is chosen to fix the gauge of the Bogomolny equation, but we
have not established this.

\section{Energy of 2-Uniton Solutions}
\label{sec:energy}
\setcounter{equation}{0}

The energy density of the field $J$ in the Ward model is
\be
\E_J = -\half\Tr((J^{-1}\pr_t J)^2 + (J^{-1}\pr_x J)^2 + (J^{-1}\pr_y J)^2) \,.
\ee
This is actually the same as in the standard Lorentz invariant sigma
model -- it is the field equation of the Ward model that breaks the
Lorentz invariance. If we use the complex coordinate $z=x+iy$ then 
\be
\E_J = -\half\Tr((J^{-1}\pr_t J)^2 
+ 4(J^{-1}\pr_z J)(J^{-1}\pr_{\bar{z}} J)) \,.
\ee
The contribution $-\half\Tr(J^{-1}\pr_t J)^2$ is the kinetic energy
density; the remaining gradient terms can be thought of as the
potential energy density. The total energy is $E=\int \E_J \, d^2x$.

Our main result is that this energy density is the same as that of a 
$\CP^3$ sigma model lump solution. In the $\CP^3$ sigma model
(Zakrzewski 1989), the
basic field is a 4-vector of complex functions in the plane
\be
V=(V_1,V_2,V_3,V_4) \,,
\ee
whose energy density and other properties are
unaffected by multiplying all components by a common function
$W$. The components of $V$ should not simultaneously vanish.
For a finite-energy multi-lump solution, all components of $V$ must be
rational functions of $z$. $V$ can then be converted to a 4-vector of
polynomials in $z$, with no common root, by multiplying through by 
the common denominator. (For an anti-lump, one takes rational functions of
$\bar{z}$ instead. There are also higher energy saddle point solutions
which depend on $z$ and $\bar{z}$, but these do not concern us here.)

The energy density in the sigma model is
\be
\label{CPEner}
\E_V = 8\,\frac{\sum_{1 \le i<j \le 4}|V_iV'_j - V_jV'_i|^2}
{(|V_1|^2 + |V_2|^2 + |V_3|^2 + |V_4|^2)^2} \,,
\ee
where $V'_i = \frac{dV_i}{dz}$. The topological charge $N$ of the lump 
is the degree of $V$, ${\rm deg} \, V$, which is the highest degree among 
the polynomials comprising $V$ 
(after clearing denominators), and the total energy $\int \E_V \, d^2x$ is
$8\pi N$. This follows from a Bogomolny argument which reduces the
energy to $8\pi$ times the integral over the plane of the pull-back 
of the 2-form generating the integer cohomology ring of the manifold $\CP^3$.

The multi-lump that corresponds to the 2-uniton solution of the Ward model
has the specific form
\be
\label{VWard}
V=(2(tf'+h), f^2, \sqrt{2}f, 1) \,.
\ee
Note that $t$ appears here explicitly as a parameter, but we calculate 
the energy density at each instant as if the multi-lump were static,
using the formula (\ref{CPEner}).

To find the total energy, we need to determine the highest degree of 
the polynomials that
occur in (\ref{VWard}) when one clears denominators. Let ${\rm deg} \,f$ 
and ${\rm deg}\,h$ denote the degrees of the rational functions $f$ and
$h$. Generically, $f=\frac{r}{s}$ and $h=\frac{u}{v}$, where $r$ and
$s$ are polynomials of degree ${\rm deg}\,f$, $u$ and $v$ are
polynomials of degree ${\rm deg}\,h$, and $s$ and $v$ have no common roots.
Then substituting in (\ref{VWard}) and multiplying by the common denominator
$s^2 v$, we find
\be
V=(2t(sr'v-rs'v)+2us^2, r^2v, \sqrt{2}rsv, s^2v) \,,
\ee
so $N=2\,{\rm deg}\,f + {\rm deg}\,h$. However there are plenty of non-generic
possibilities. For example, if $f$ and $h$ are polynomials, then $N$
is the greater of $2\,{\rm deg}\,f$ and ${\rm deg}\,h$. 
Generically, $N$ is also the number of zeros of $tf'+h$, but again there
are non-generic examples where this is not the case.

Now consider the energy density for $V$ of the form
(\ref{VWard}). Since time derivatives play no role here, let us
denote $2(tf'+h)$ by $g$. One finds after some algebra that
\be
\label{VEner}
\E_V = \frac{8|(1+|f|^2)g' - 2g\bar{f} f'|^2 + 16|gf'|^2 +16(1+|f|^2)^2 |f'|^2}
{(|g|^2 + (1+|f|^2)^2)^2} \,. 
\ee

We now wish to demonstrate that $\E_J=\E_V$. Let us consider the
kinetic contribution to $\E_J$ first. Note that $J=AB$ (the product of the two
uniton factors in (\ref{2Unit})), and that $A^{-1} =
A$ and $B^{-1} = B$. Note also that $B$ is independent of $t$. Hence 
$\Tr(J^{-1}\pr_t J)^2 = -\Tr(\pr_t A \pr_t A)$. Further
simplification is possible by expressing $q_1$ as a linear combination
of the everywhere orthonormal, time independent row vectors
$(1+|f|^2)^{-\half} (1,f)$ and $(1+|f|^2)^{-\half}
(\bar{f},-1)$. Their coefficients are, respectively,
$(1+|f|^2)^{\half}$ and $-2i(1+|f|^2)^{-\half} g$. After a
modest calculation involving these coefficients, and noting that
the time derivative of $g$ is $2f'$, one finds that
\be
-\half\Tr(J^{-1}\pr_t J)^2 = \frac{16(1+|f|^2)^2 |f'|^2}
{(|g|^2 + (1+|f|^2)^2)^2} \,,
\ee
which is identical to the contribution to $\E_V$ from the third term in the
numerator of (\ref{VEner}). A rather more involved calculation, using 
MAPLE, leads to the result that the gradient energy is
\be
-2\Tr(J^{-1}\pr_z J)(J^{-1}\pr_{\bar{z}} J) = 
\frac{8|(1+|f|^2)g' - 2g\bar{f}f'|^2 + 16|gf'|^2}{(|g|^2 + (1+|f|^2)^2)^2},
\ee
which is identical to the contribution of the first and second terms
in the numerator of (\ref{VEner}).
In total, we see that $\E_J=\E_V$, and note that in $\E_J$ we include
time derivatives, but in $\E_V$ we do not.

Let us illustrate these formulae with a pair of examples. The first is
the two-soliton solution, with $f(z)=z$ and $h(z)=z^2$. The energy
density is
\be
\E = 16\,\frac{5r^4 + 10r^2 + 1 + 4t^2(2r^2+1) - 8t(x^2-y^2)}
{(5r^4 + 2r^2 + 1 + 4t^2 + 8t(x^2-y^2))^2},
\ee
where $r^2=x^2+y^2$. The energy is peaked at two points on the
$x$-axis for $t<0$, forms a ring around the origin at $t=0$, and is
peaked at two points on the $y$-axis for $t>0$. See Ward (1995) for
figures showing the energy distribution at different times. The total
energy is $16\pi$, because the highest degree of the polynomials in
$V$ is 2. This is also easily verified by direct integration of the circularly
symmetric energy density at $t=0$, and at other times follows from
energy conservation.

A second example is $f(z)=z$ and $h(z)=z^3$. Here the energy density is
\be
\E = 16\,\frac{2r^8 + 16r^6 + 19r^4 + 2r^2 + 1 + 4t^2(2r^2+1) - 
8t(r^2+2)x(x^2-3y^2)}
{\left(4r^6 + r^4 + 2r^2 + 1 + 4t^2 + 8tx(x^2-3y^2)\right)^2},
\ee
(which simplifies and corrects formula (21) of Ioannidou (1996)). This
energy density exhibits the scattering of three lumps with triangular
symmetry, whose locations are approximately at the
zeros of $z^3+t$ (see Ioannidou (1996) for figures). The total 
energy is $24\pi$.

\begin{figure}[b]
\hskip .2cm
\put(59,125){$t=-0.5$} 
\hskip 0.6cm
\epsfxsize=8cm\epsfysize=7cm\epsffile{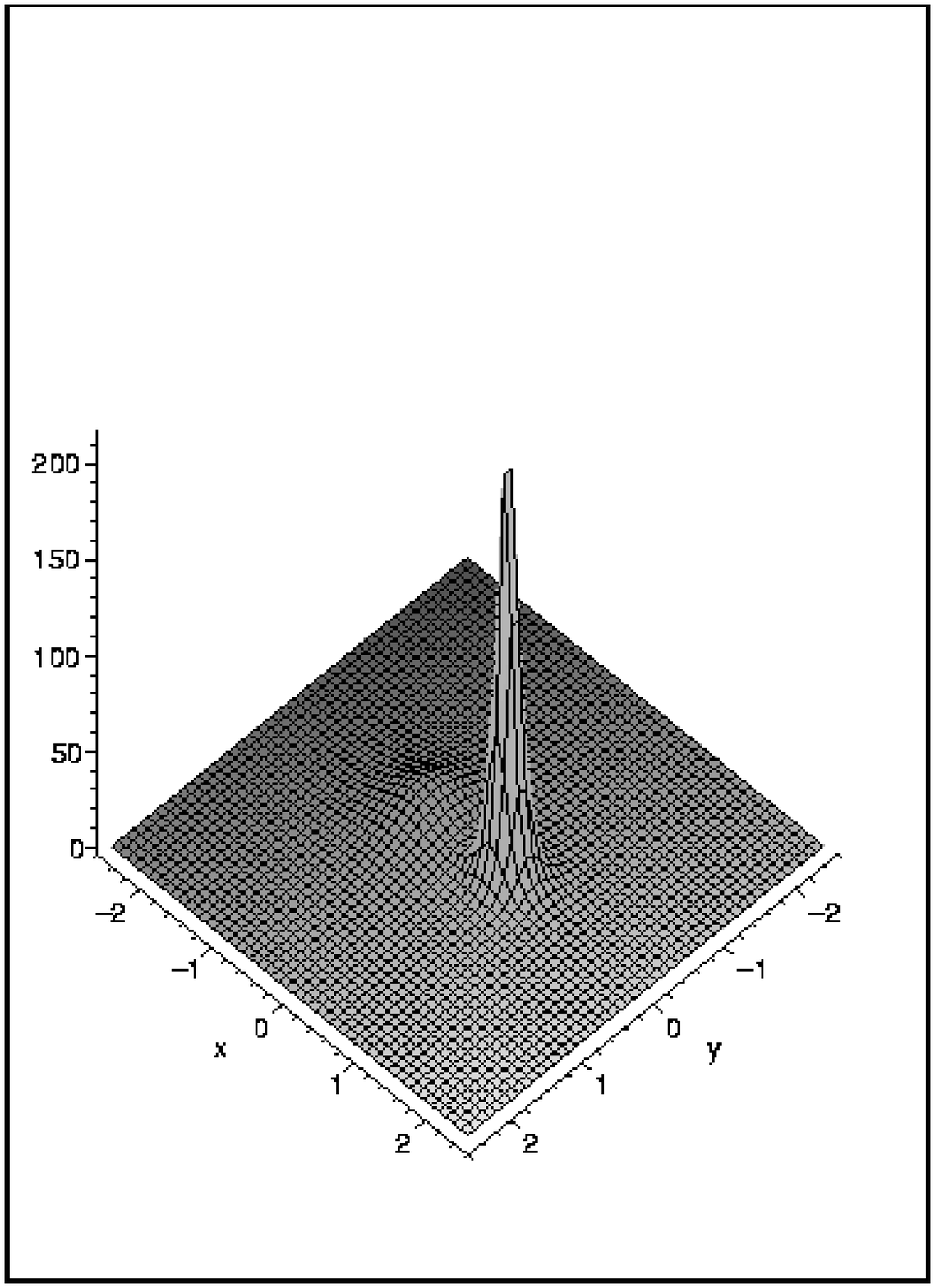}
\hfill 
\hskip 0.1cm
\put(49,125){$t=0$}
\epsfxsize=8cm\epsfysize=7cm\epsffile{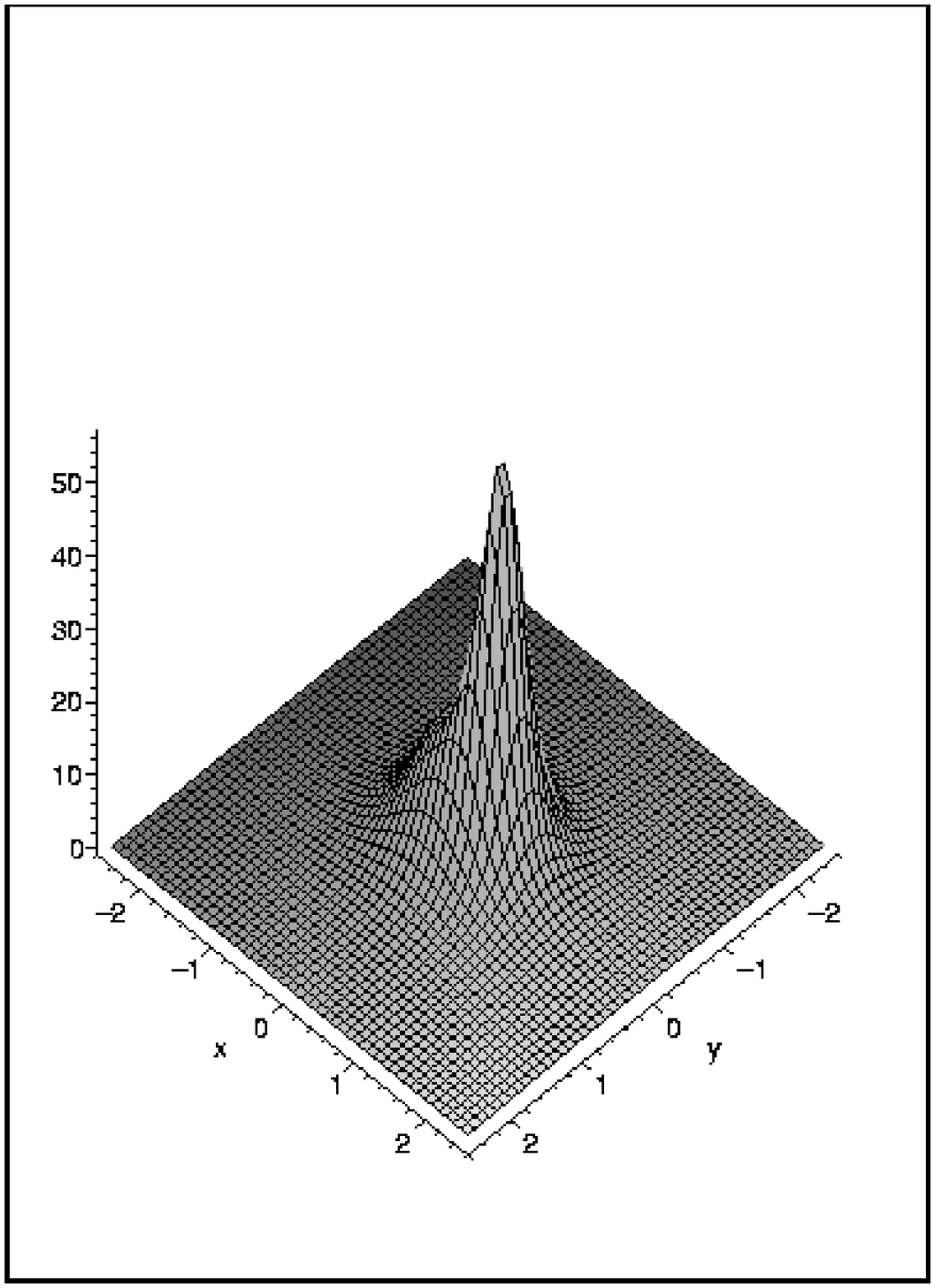}
\vskip 0.25cm
\hskip 1.2cm
\put(59,125){$t=1$} 
\hskip 0.6cm
\epsfxsize=8cm\epsfysize=7cm\epsffile{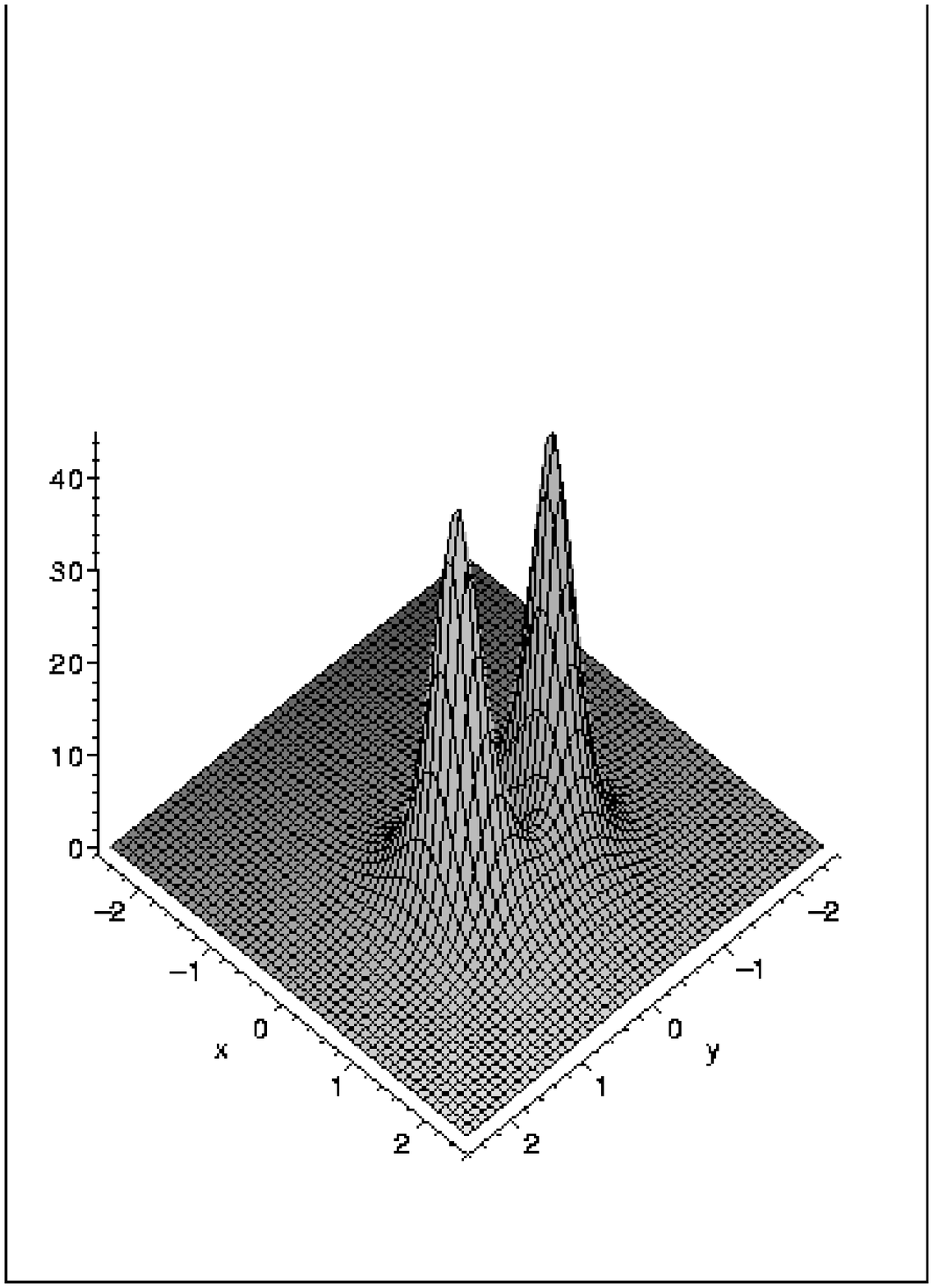}
\hfill 
\hskip 0.1cm
\put(49,125){$t=3.5$} 
\epsfxsize=8cm\epsfysize=7cm\epsffile{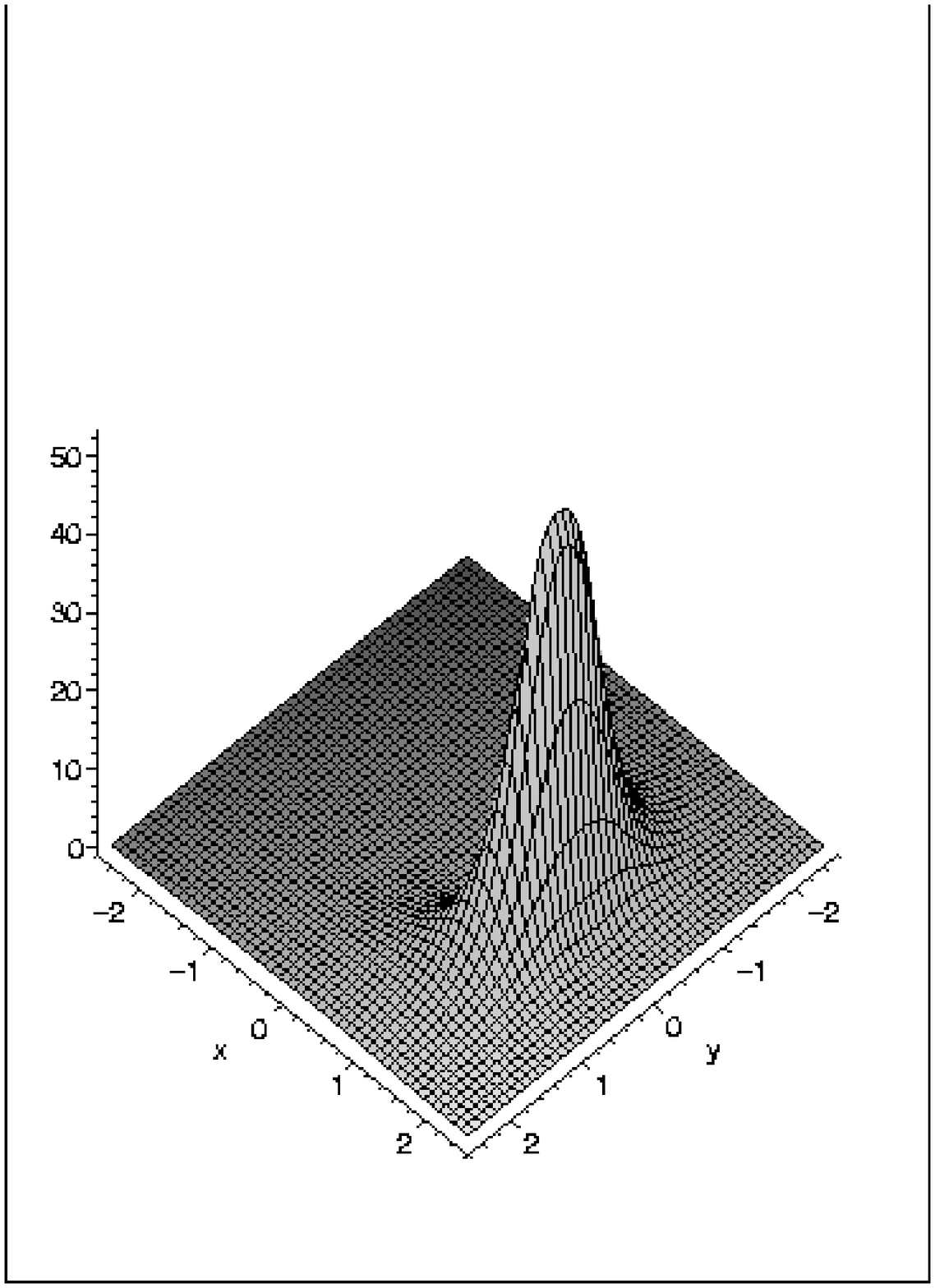}
\vskip 0.25cm
\hskip 1.2cm
\put(59,125){$t=4$} 
\hskip 0.6cm
\epsfxsize=8cm\epsfysize=7cm\epsffile{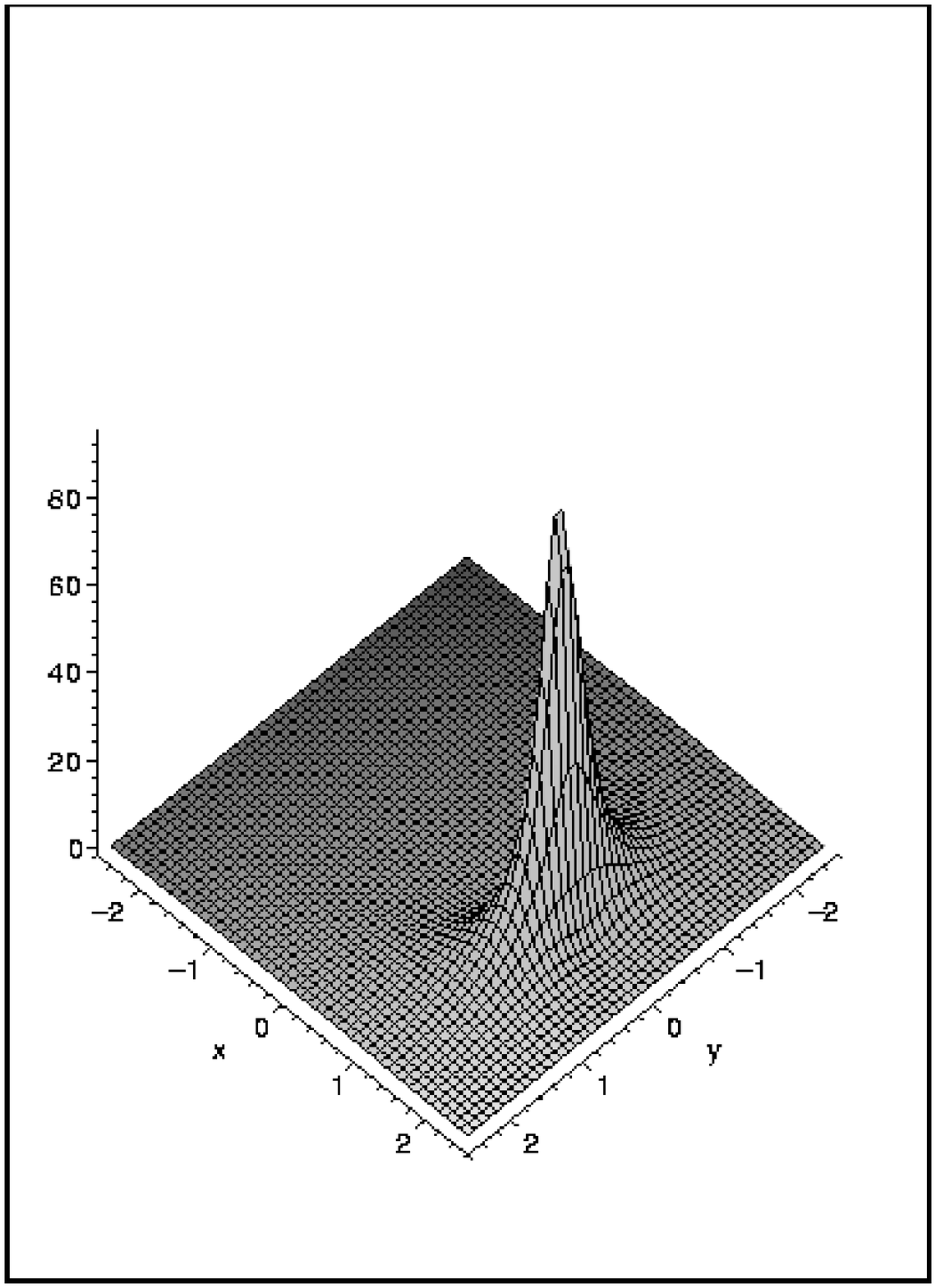}
\hfill 
\hskip 0.1cm
\put(49,125){$t=5$} 
\epsfxsize=8cm\epsfysize=7cm\epsffile{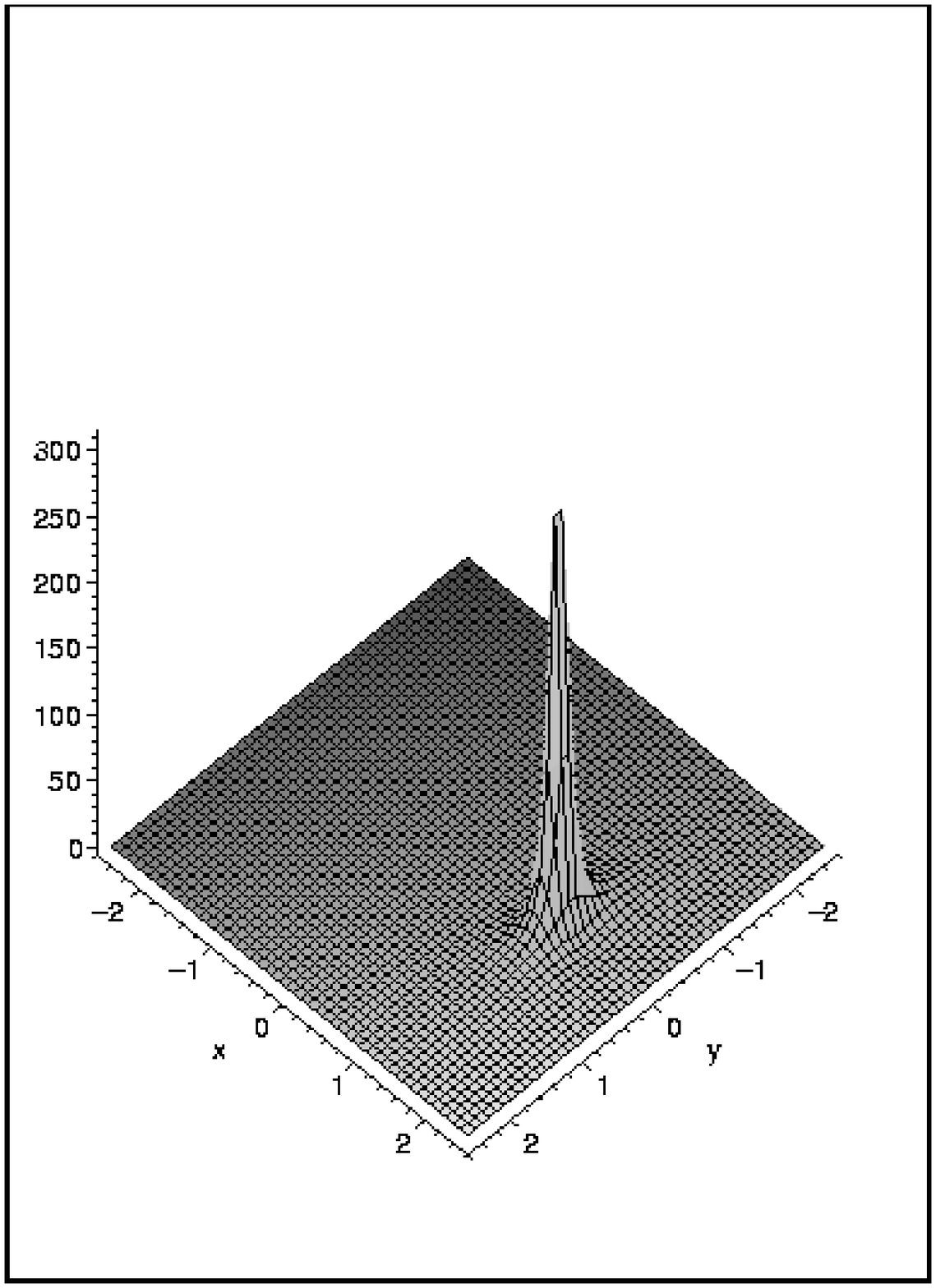}
\vskip -.5cm
\caption{The soliton configuration at different times for the generic 
example $f=1/z$ and $h=1/(z-1)$.}
\label{Fig1}
\end{figure}

We have also investigated other cases where $f$ and $h$ are rational
functions, but not polynomials. Figure \ref{Fig1} presents an example
of generic type, where $f=1/z$ and $h=1/(z-1)$. The expression for the 
energy density is complicated, but it is easily seen that ${\rm deg}
\, V = 3$, implying that the total energy is $24\pi$ (as confirmed 
numerically). In addition, the motion of the configuration is 
interesting as seen from the zeroes of $tf' + h$, that is, the
zeroes of $z^2 - tz + t$. These show the locations of a pair of
soliton structures whose motion is not so simple. For $t<0$ and $t>4$ 
both zeros are real, but for $0<t<4$ the zeros are a complex conjugate pair.
At $t=0$ and $t=4$ there are repeated zeros at $z=0$ and $z=2$, respectively.
In Figure \ref{Fig1} we can see solitons approaching along the
$x$-axis at early times, colliding at $t=0$ and scattering at
right angles, then moving apart on circular orbits and colliding again 
at $t=4$, and finally separating along the $x$-axis at late times. 

\begin{figure}[b]
\hskip .2cm
\put(49,125){$t=-2$} 
\hskip 0.6cm
\epsfxsize=8cm\epsfysize=8cm\epsffile{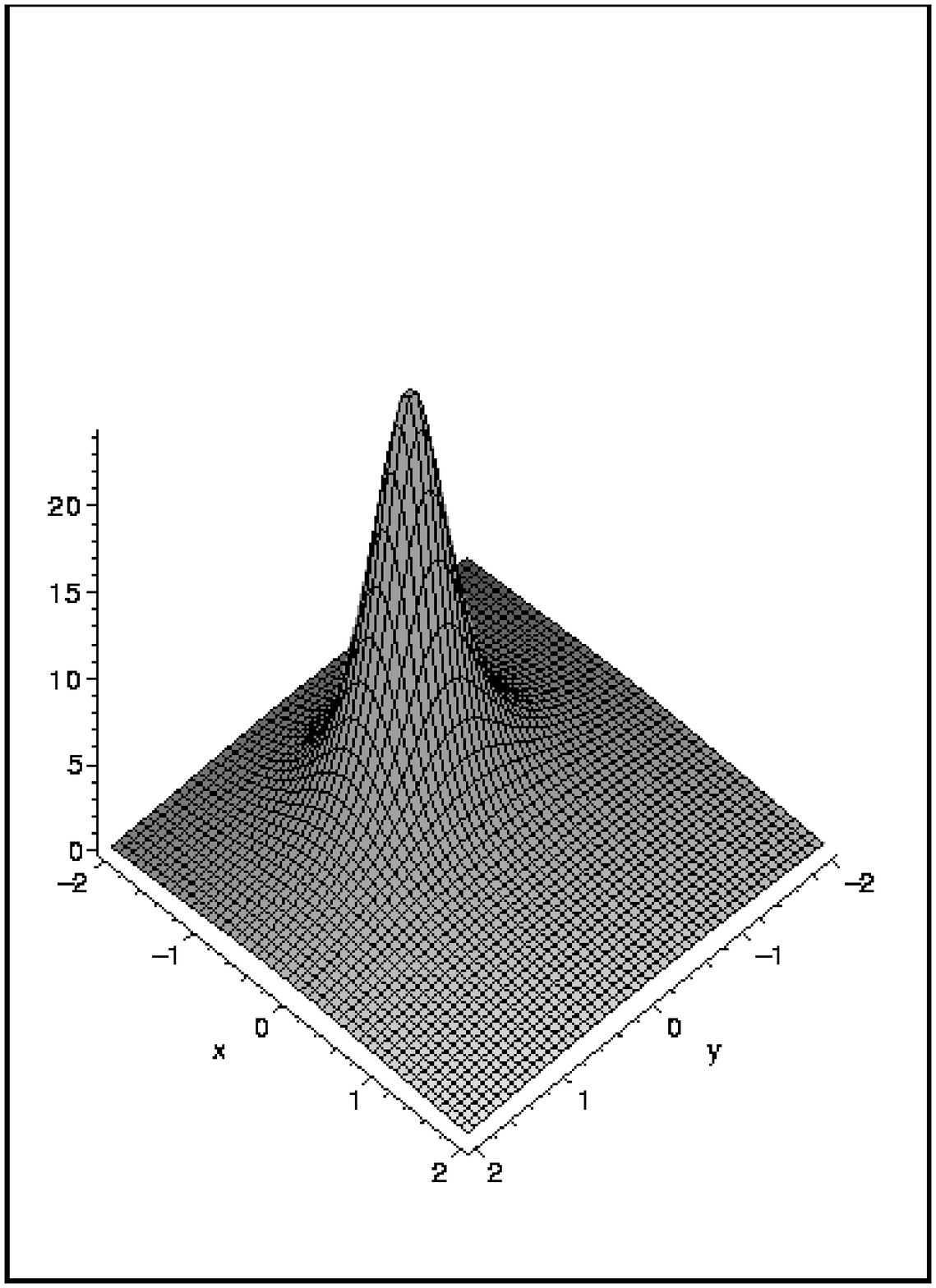}
\hfill 
\hskip 0.1cm
\put(49,125){$t=-1$}
\epsfxsize=8cm\epsfysize=8cm\epsffile{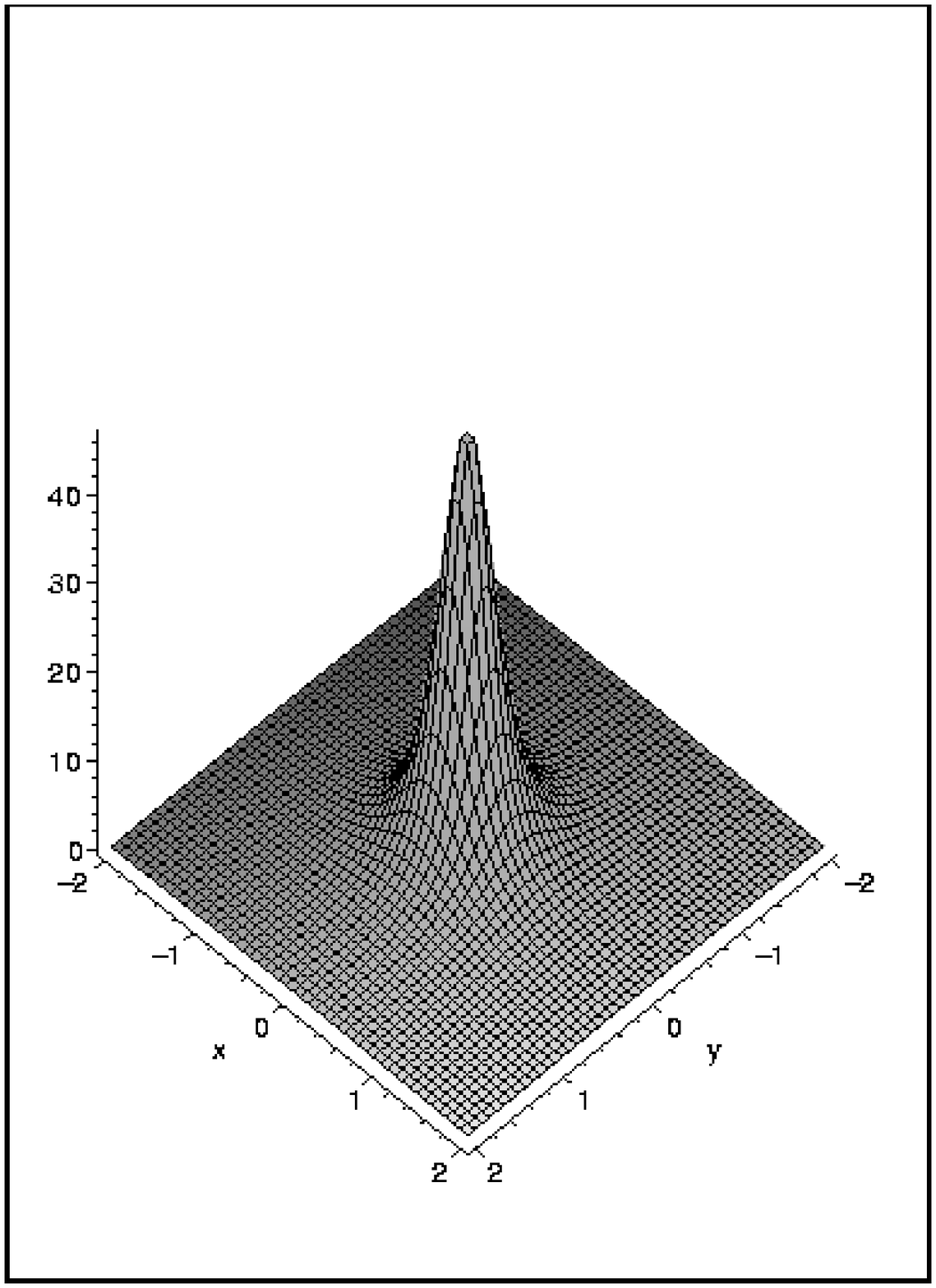}
\vskip -1cm
\vskip 1.25cm
\hskip 1.2cm
\put(100,120){$t=0$} 
\hskip 0.6cm
\epsfxsize=8cm\epsfysize=8cm\epsffile{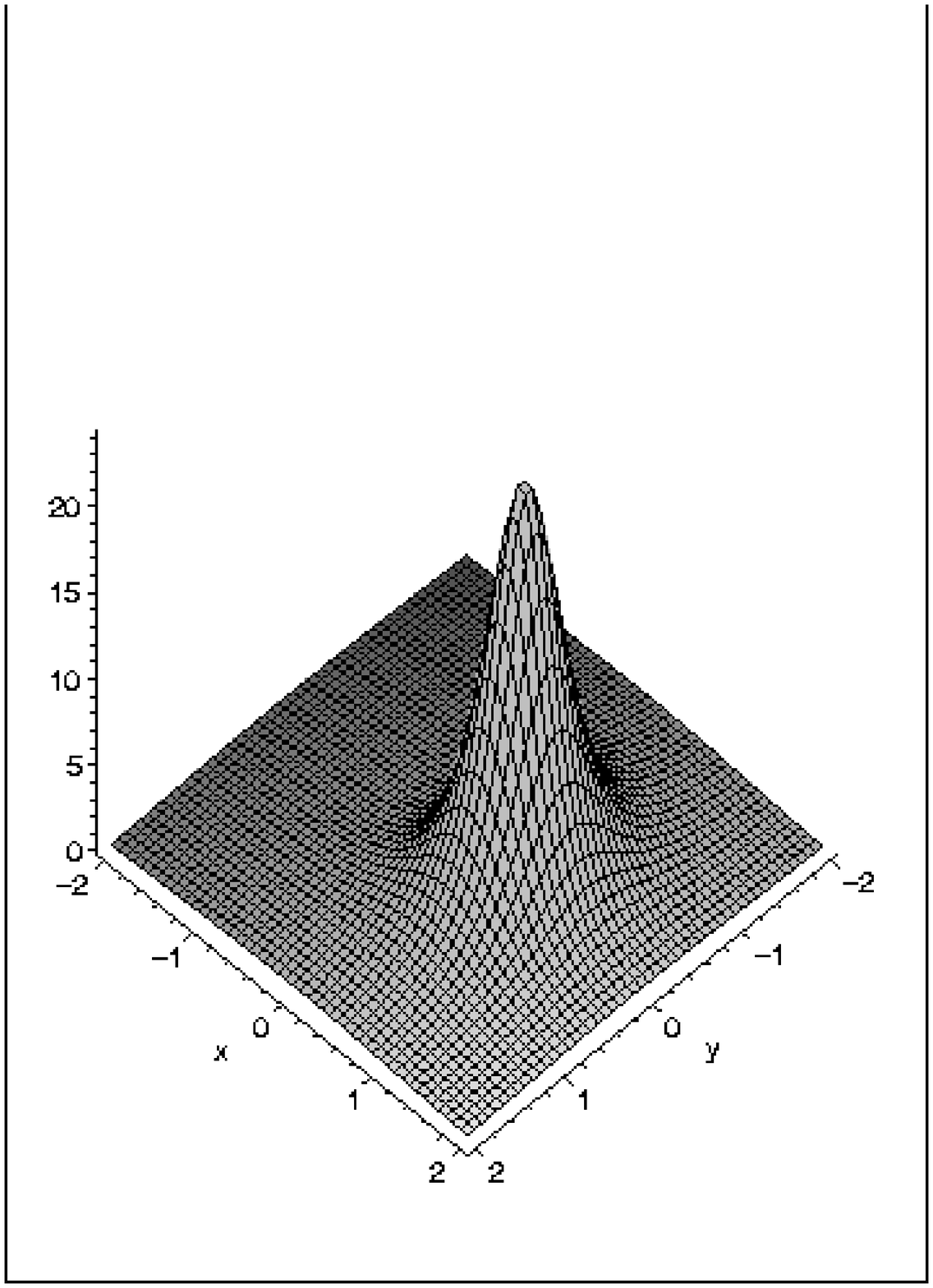}
\vskip -.5cm
\caption{The soliton configuration at different times for the non-generic 
example $f=1/z$ and $h=(z-1)/z^2$.}
\label{Fig2}
\end{figure}

An example of the non-generic type (because the denominator of $h$ is
the square of the denominator of $f$) is $f(z)=1/z$ and
$h(z)=(z-1)/z^2$. Here, the energy density is
\be
\E=16\frac{3r^4 + 2r^2 + 3 + 8\tau^2 r^2 + 4\tau^2 - 8\tau r^2 x}
{\left(r^4 + 6r^2 + 1 + 4\tau^2 - 8\tau x \right)^2} \,,
\ee
where $\tau = t+1$. This is symmetric under the simultaneous
reflections $x \to -x$ and $\tau \to -\tau$. Figure \ref{Fig2}
presents the evolution of the configuration for times around $t=-1$.
The total energy is $16\pi$.

\section{Generalisations}
\label{sec:gen}
\setcounter{equation}{0}

More complicated multi-pole solutions of the Ward model can be
constructed. For a recent, general analysis of these, see
Dai \& Terng (2004). A triple-pole example, for which $J$ has a 3-uniton
form, is presented in Ioannidou (1996). In this, and a few further
examples, we have verified that the total energy is an integer
multiple of $8\pi$. We suspect that the instantaneous
energy density can again be identified with that of a static
sigma model solution, but now for $\CP^5$. Clarification of this, and its
generalisation, would be desirable.

\vskip 20pt
\centerline{\bf{Acknowledgement}}
\vspace{.25cm}
\noindent
N.S.M. thanks the Mathematics Division of the Aristotle University of 
Thessaloniki for hospitality.

\newpage
\centerline{\bf{References}}
\vspace{.25cm}
\noindent 
Dai, B. \& Terng, C.-L. 2004 
B\"acklund transformations, Ward solitons, and unitons. 
Preprint math.DG/0405363.

\noindent
Fokas, A. S. \& Ioannidou, T. A. 2001 The inverse spectral
theory for the Ward equation and for the 2+1 chiral model. 
{\sl Commun. Appl. Anal.} {\bf 5}, 235--246.

\noindent
Ioannidou, T. 1996 Soliton solutions and nontrivial
scattering in an integrable chiral model in (2+1) dimensions. 
{\sl J. Math. Phys.} {\bf 37}, 3422--3441.

\noindent
Ioannidou, T. \& Ward, R. S. 1995 Conserved quantities for
integrable chiral equations in 2+1 dimensions. {\sl Phys. Lett. A} 
{\bf 208}, 209--213.

\noindent
Ioannidou, T. \& Zakrzewski, W. J. 1998 Solutions of the
modified chiral model in (2+1) dimensions. 
{\sl J. Math. Phys.} {\bf 39}, 2693--2701.

\noindent
Manton, N. S. \& Sutcliffe, P. M. 2004 
{\sl Topological solitons}. Cambridge University Press.

\noindent
Uhlenbeck, K. 1989 Harmonic maps into Lie groups
(Classical solutions of the chiral model). 
{\sl J. Diff. Geom.} {\bf 30}, 1--50.

\noindent
Ward, R. S. 1988 Soliton solutions in an integrable
chiral model in 2+1 dimensions. {\sl J. Math. Phys.} {\bf 29}, 386--389. 
\vspace{.1cm}

\noindent
Ward, R. S. 1995 Nontrivial scattering of localized
solitons in a (2+1)-dimensional integrable system. 
{\sl Phys. Lett. A} {\bf 208}, 203--208.

\noindent
Zakrzewski, W. J. 1989 {\sl Low dimensional sigma models}. 
Institute of Physics Publishing.


\end{document}